\documentclass[twocolumn,prl,aps,showpacs,superscriptaddress]{revtex4}

\usepackage{mathptmx}
\usepackage{subfigure}
\usepackage{psfrag,graphicx}
\usepackage{dcolumn}
\usepackage{amsmath,amssymb}
\usepackage{bm}
\usepackage{color}
\usepackage{latexsym}
\usepackage{epstopdf}
\usepackage{color}
\usepackage[english]{babel}
\usepackage{latexsym}
\usepackage{psfrag,graphicx}
\usepackage{subfigure}
\usepackage{amsmath}
\usepackage{amssymb}
\usepackage{amsfonts}
\usepackage{bm}
\usepackage{natbib}
\usepackage{epstopdf}
\usepackage{appendix}

\definecolor{mygrey}{gray}{0.35}
\definecolor{myblue}{rgb}{0.2,0.2,0.8}
\definecolor{myzard}{cmyk}{0,0,0.05,0}
\definecolor{mywhite}{rgb}{1,1,1}
\definecolor{mywhite}{rgb}{1,1,1}
\definecolor{myred}{rgb}{1,0.,0.3}

\usepackage[colorlinks=true,citecolor=myblue,linkcolor=myred]{hyperref}

\def\ba{\begin{align}}
\def\enda{\end{align}}
\def\bi{\begin{itemize}}
\def\ei{\end{itemize}}

\def\be{\begin{equation}}
\def\ee{\end{equation}}
\def\bea{\begin{eqnarray}}
\def\eea{\end{eqnarray}}
\def\bse{\begin{subequations}}
\def\ese{\end{subequations}}

\def\sec#1{\textbf{#1.}}
\def\subsec#1{\textit{#1.}}

\begin{document}
\title{Spontaneous-symmetry-breaking assisted quantum sensors}
\author{Peter A. Ivanov}
\affiliation{Department of Physics, Sofia University, James Bourchier 5 blvd, 1164 Sofia, Bulgaria}
\author{Kilian Singer}
\affiliation{QUANTUM, Institut f\"ur Physik, Universit\"at Mainz, D-55128, Germany}
\author{Nikolay V. Vitanov}
\affiliation{Department of Physics, Sofia University, James Bourchier 5 blvd, 1164 Sofia, Bulgaria}
\author{Diego Porras}
\affiliation{Department of Physics and Astronomy, University of Sussex, Falmer, Brighton BN1 9QH, UK}

\begin{abstract}
We propose a quantum sensing scheme for measuring weak forces based on a symmetry-breaking adiabatic transition in the quantum Rabi model.
We show that the system described by the Rabi Hamiltonian can serve as a sensor for extremely weak forces with sensitivity beyond the yN $/\sqrt{\text{Hz}}$ range.
We propose an implementation of this sensing protocol using a single trapped ion.
A major advantage of our scheme is that the force detection is performed by projective measurement of the population of the spin states at the end of the transition, instead of the far slower phonon number measurement used hitherto.
\end{abstract}

\pacs{
03.67.Ac, %Quantum computation architectures and implementations
03.67.Bg,
03.67.Lx,
42.50.Dv %Coherent control of atomic interactions with photons
}
\maketitle

%%%%%%%%%%%%%%%%%%%%%%%%%%%%%%%%%%%%%%%%%%%%%%%%%%%%%%%%%%%%%%%%%%%%%%%%%%%
%%%%%%%%%%%%%%%%%%%%%%%%%%%%%%%%%%%%%%%%%%%%%%%%%%%%%%%%%%%%%%%%%%%%%%%%%%%
%%%%%%%%%%%%%%%%%%%%%%%%%%%%%%%%%%%%%%%%%%%%%%%%%%%%%%%%%%%%%%%%%%%%%%%%%%%
%========================================================================
%========================================================================
\sec{Introduction}  Using nanoscale mechanical oscillators as detectors of extremely weak forces has attracted considerable experimental interest \cite{Mamin2001}. Such systems allow to measure forces with sensitivity below the attonewton range which is beneficial for a broad range of applications. For example, a force detector with a nanomechanical oscillator coupled to a microwave cavity can reach sensitivity below one aN ($10^{-18}$ N) per $\sqrt{\rm{Hz}}$ \cite{Teufel2009}. Other sensors use mechanical oscillators made of carbon nanotubes for force detection with sensitivity in the zN ($10^{-21}$ N) per $\sqrt{\rm{Hz}}$ range \cite{Moser2013}. Recently, the detection of ultra-weak forces as small as $5$ yN ($10^{-24}$ N) was experimentally demonstrated using injection locked ions \cite{Knunz2010}. Force measurement in an ensemble of ions in a Penning trap uses Doppler velocimetry technique to detect force with sensitivity of $170$ yN$/\sqrt{\rm{Hz}}$ \cite{Biercuk2010}.
Another approach uses high-precision ion position determination to measure light pressure forces \cite{Gloger2015}.
%This principle is used as a basis for several kinds of quantum sensors in which the external applied field is detected as a phase shift.
In all cases the force sensing based on mechanical oscillators is carried out by converting the force into a displacement that is measured by electrical or optical means.

In this Letter, we introduce a different sensing protocol, which uses a system described by the quantum Rabi (QR) model as a probe that is sensitive to extremely weak forces. The QR model consists of a single bosonic mode and an effective spin system which interact via dipolar coupling. We show that the effect of symmetry breaking of the underlying parity symmetry in the QR model due to the presence of external perturbation can be used in an efficient way for detection of classical oscillating forces. Our scheme relies on the adiabatic evolution of the ground state of the QR model into the Schr\"odinger cat state, where the relevant force information is mapped in the respective probability amplitudes.
The force sensing is performed simply by measuring the spin populations.
{Therefore our protocol, which demands a single population measurement, is considerably faster than previous protocols based on the detection of the motional degree of freedom via Rabi oscillations.}

We consider a particular implementation of our sensing scheme using a coherently manipulated single trapped ion.
The scheme, however, can be realized with various quantum optical systems such as nitrogen vacancy centers in diamond and superconducting qubits inside a microwave cavity \cite{Niemczyk2010,Ballester2012,Zou2014}.
We show that with the current ion trap technologies force sensitivity below one yN$/\sqrt{\rm{Hz}}$ can be achieved.
In addition, we show that our method can be applied for detection of spin-dependent forces which are created in magnetic-field gradients or Stark-shift gradients.
Hence our method can be used for studying magnetic dipole moments of atomic or molecular ions.

%The paper is organized as follows.
%In Sec. \ref{AQM} we introduce the sensing protocol based on the symmetry-breaking adiabatic transition in the QR model.
%In Sec. \ref{ES} we describe the sensing scheme for detection of weak forces using single trapped ion.
%Finally, in Sec. \ref{C} we summarize our findings.

\sec{Adiabatic quantum metrology using the quantum Rabi model}
Our system consists of a two-level atom with states $\left|\uparrow\right\rangle$ and $\left|\downarrow\right\rangle$ coupled to a single bosonic mode described by the quantum Rabi model,
\begin{equation}
\hat{H}_{\rm R}=\hbar\omega \hat{a}^{\dag}\hat{a}+\frac{\hbar\Omega_{y}(t)}{2}\sigma_{y}+\hbar g\sigma_{x}(\hat{a}^{\dag}+\hat{a}).\label{Dicke}
\end{equation}
Here $\hat{a}^{\dag}$ and $\hat{a}$ are the creation and annihilation operators of bosonic excitation with frequency $\omega$ and $\sigma_{\beta}$ ($\beta=x,y,z$) are the respective Pauli matrices.
The time-dependent Rabi frequency of the transverse field is given by $\Omega_{y}(t)$ and $g$ is the spin-boson coupling.
Recently, it was shown that the Rabi model permits exact integrability \cite{Braak2011}.

The quantum Rabi Hamiltonian (\ref{Dicke}) possesses a discrete symmetry revealed by the parity transformation $\hat{a}\rightarrow - \hat{a}$, $\sigma_{y}\rightarrow \sigma_{y}$ and $\sigma_{x}\rightarrow -\sigma_{x}$.
In the following we consider the QR model in the ultrastrong coupling regime $g/\omega<1$ and study the effect of a small perturbation term $\hat{H}_{\rm pert}$, which breaks the underlying parity symmetry of the model. The total Hamiltonian including the perturbation term becomes
\begin{equation}
\hat{H}=\hat{H}_{\rm R}+\hat{H}_{\rm pert}.\label{H}
\end{equation}
As we will see, the symmetry-breaking process allows us to estimate the perturbation term very accurately.
\begin{figure}
\includegraphics[width=0.45\textwidth]{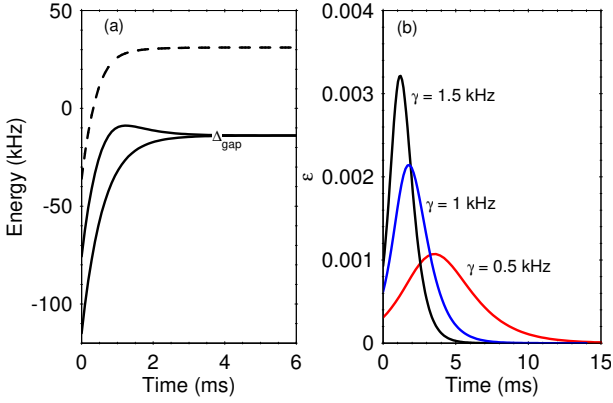}
\caption{(Color online) (a) Low-energy spectrum of the quantum Rabi model as a function of time $t$. The two lowest lying states are separated by energy splitting $\Delta_{\rm gap}$. At the initial moment $t=0$ the system is prepared in the ground-state with $\Omega_{y}(0)\gg g$ and then it evolves into the Schr\"odinger cat state (\ref{SC}). The energy difference between the third excited energy (dashed line) and the ground-state energy is $\Delta_{\rm ge}$. We assume $\gamma=1.5$ kHz, $g=25$ kHz, $\omega=45$ kHz and $\Omega_{y}(0)=225$ kHz. (b) Adiabatic parameter $\epsilon$ versus time $t$ for different values of $\gamma$.}
\label{adcond}
\end{figure}

The measurement protocol for $\hat{H}_{\rm pert}$ starts by preparing the system in the ground-state of the unperturbed Hamiltonian $\hat{H}_{\rm R}$ in the limit $\Omega_{y}(0)\gg g$, such that $\left|\psi_{\rm g}(0)\right\rangle=\left|-\right\rangle_{y}\left|0\right\rangle$ where $\left|-\right\rangle_{y}=(\left|\uparrow\right\rangle-\rm{i}\left|\downarrow\right\rangle)/\sqrt{2}$ and $\left|n\right\rangle$ is the Fock state of the bosonic mode with occupation number $n$. Then we adiabatically decrease the transverse field $\Omega_{y}(t)$ in time such that the system evolves into the Schr\"odinger cat state
\begin{equation}
|\psi_{\rm g}(t)\rangle=c_{+}(t)|\psi_{+}\rangle+c_{-}(t)|\psi_{-}\rangle,\label{SC}
\end{equation}
where $c_{\pm}(t)$ are the respective probability amplitudes. Here $\left|\psi_{+}\right\rangle=\left|+\right\rangle_{x}\left|\alpha\right\rangle$ and $\left|\psi_{-}\right\rangle=\left|-\right\rangle_{x}\left|-\alpha\right\rangle$ form the ground-state multiplet with $\left|\pm\right\rangle_{x}=(\left|\uparrow\right\rangle\pm\left|\downarrow\right\rangle)/\sqrt{2}$ and $|\alpha\rangle$ stands for a coherent state with amplitude $\alpha=-g/\omega$. State (\ref{SC}) implies that for $\hat{H}_{\rm pert}=0$ the parity symmetry is preserved by creating an entangled state with equal probabilities, $c_{\pm}=\pm1/\sqrt{2}$. The effect of the perturbation is to break the parity symmetry of $\hat{H}_{\rm R}$ by creating a ground-state wavefunction (\ref{SC}) with unequal probability amplitudes, $|c_{+}|^{2}\neq|c_{-}|^{2}$. By measuring the respective probabilities at the end of the process one can estimate the unknown perturbation.

In order to describe the creation of the symmetry-broken ground state we represent the Hamiltonian \eqref{H} within the ground-state multiplet, which leads to an effective two-level problem with the Hamiltonian \cite{Ivanov2013}
\begin{equation}
H_{\rm eff} =\left[
\begin{array}{cc}
\left\langle\psi_{+}\right|\hat{H}_{\rm pert}\left|\psi_{+}\right\rangle & \hbar\Delta_{\rm gap}/2 \\
\hbar\Delta_{\rm gap}/2 & \left\langle\psi_{-}\right|\hat{H}_{\rm pert}\left|\psi_{-}\right\rangle  \\%
\end{array}%
\right].\label{Heff1}
\end{equation}
Here $\Delta_{\rm gap}$ is the energy gap of the ground-state multiplet which scales with the transverse field as $\Delta_{\rm gap}\sim \Omega_{y}$ (see, Fig. \ref{adcond}a) \cite{Ivanov2013}. Hereafter we assume an exponential decay of the transverse field $\Omega_{y}(t)=\Omega_{y}(0)e^{-\gamma t}$ with a characteristic slope $\gamma$ which implies that $\Delta_{\rm gap}\sim e^{-\gamma t}$. The adiabaticity of the process is characterized by the condition $\epsilon=|\left\langle \psi_{\rm g}\right|d/dt\left|\psi_{\rm e}\right\rangle/\Delta_{\rm ge}| \ll 1$, which requires the coupling between the ground-state $\left|\psi_{\rm g}\right\rangle$ to the first excited-state $\left|\psi_{\rm e}\right\rangle$ to be much smaller that the energy gap between them $\Delta_{\rm ge}$ at any instant of time.
In Fig. \ref{adcond}(b) we show the adiabatic parameter $\epsilon$ during the creation of the Schr\"odinger cat state (\ref{SC}) for various $\gamma$.
We observe that $\epsilon\ll 1$ and the non-adiabatic transitions are weaker for lower $\gamma$.

At the final stage the externally applied perturbation is detected by measuring the expectation value of $\sigma_{x}$.
An analytical expression for the measured signal can be derived by solving the quantum evolution of a two-level system with the Hamiltonian \eqref{Heff1} for the specific time-dependence of $\Omega_{y}(t)$.

\sec{Sensing weak forces and displacements}
In the following we consider a harmonic oscillator represented by a single trapped ion with mass $m$ confined in a Paul trap with an axial trap frequency $\omega_z$.
We assume that the effective spin system of the ion is implemented by two metastable atomic levels $\left|\uparrow\right\rangle$ and $\left|\downarrow\right\rangle$ with a transition frequency $\omega_{0}$.
We describe the small axial vibrations of the ion by the following motional Hamiltonian
\begin{equation}
\hat{H}_{\rm m}=\hbar\omega\hat{a}^{\dag}\hat{a},\quad \hat{z}=z_{0}(\hat{a}^{\dag}+\hat{a}),
\end{equation}
where $\hat{a}^{\dag}$ and $\hat{a}$ are the respective phonon creation and annihilation operators and $z_{0}=\sqrt{\hbar/2 m\omega_{z}}$ is the spread of the oscillator ground-state wave function \cite{Schneider2012}.

\subsec{Electric field sensing}
The ability to control the motional and internal states with high accuracy makes the trapped ions a formidable experimental tool for electric-field sensing \cite{Biercuk2010,Gloger2015,Maiwald2009}. In contrast to the conventional methods which rely only on the detection of the motional degrees of freedom \cite{Maiwald2009,Munro2002}, here the relevant information is transferred directly into the spin degrees of freedom due to the use of the symmetry-breaking adiabatic transition. In the following, we assume that a classical oscillating force $F(t)=F_{\rm d}\cos(\omega_{\rm d}t)$ with an amplitude $F_{\rm d}$ --- the parameter we wish to estimate --- and frequency $\omega_{\rm d}=\omega_{z}-\omega$ shifted from the axial trap frequency $\omega_{z}$ by a small detuning $\omega$ ($\omega_{z}\gg \omega$) is applied to the ion.
The action of the force is to displace the motional amplitude of the ion's vibrational oscillator described by
\begin{equation}
\hat{H}_{F}=F(t)\hat{z}(t)=\frac{z_{0}F_{\rm d}}{2}(\hat{a}^{\dag}e^{{\rm i}\omega t}+\hat{a}e^{-{\rm i}\omega t}),
\end{equation}
where we have neglected the fast-rotating terms. In order to implement the spin-boson term in the quantum Rabi Hamiltonian \eqref{H} we assume that the ion is simultaneously addressed by bichromatic laser fields in a Raman configuration with a wave vector difference $\Delta \vec{k}$ along the $z$ direction, which induces a transition between the spin states via an auxiliary excited state.
By setting the laser frequency beatnotes $\omega_{\rm r}=\omega_{0}-\omega_{z}+\omega$ and $\omega_{\rm b}=\omega_{0}+\omega_{z}-\omega$ close to the red- and blue-sideband transitions of the vibrational mode $\omega_{z}$, the resulting Hamiltonian in the Lamb-Dicke limit ($\eta \ll 1$) becomes \cite{Schneider2012,Lee2005}
\begin{figure}
\includegraphics[width=0.45\textwidth]{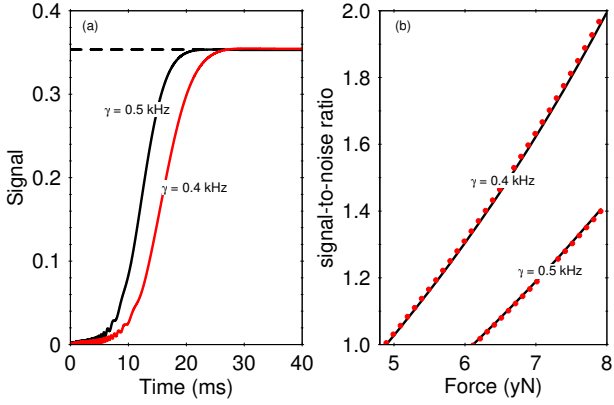}
\caption{(Color online) (a) Time-evolution of the expectation value of $\sigma_{x}$ (solid lines) for minimal detectable force (\ref{force_min}) and different $\gamma$. The dashed lines correspond to the asymptotic solution \eqref{s}. We assume q single $^{24}$Mg$^+$ trapped ion with an axial trap frequency $\omega_{z}=6.3$ MHz. The other parameters are set to $g=25$ kHz, $\omega=150$ kHz, $\Omega_{y}(0)=225$ kHz. (b) Signal-to-noise ratio versus $F_{\rm d}$. We compare the numerical solution of the time-dependent Schr\"odinger equation with the Hamiltonian \eqref{H} (dots) with the asymptotic expression given by ${\rm SNR}=\sinh(\pi g z_{0}F_{\rm d}/(\hbar \gamma\omega))$ (solid lines).}
\label{fig1}
\end{figure}
\begin{equation}
\hat{H}_{\rm s-b}=\hbar g\sigma_{x}(\hat{a}^{\dag}e^{{\rm i}\omega t}+\hat{a}e^{-{\rm i}\omega t}),\label{H_sb}
\end{equation}
where $g=\Omega\eta$ is the spin-phonon coupling with $\Omega$ being the two-photon Rabi frequency and $\eta$ stands for the Lamb-Dicke parameter.
The transverse field in Eq.~\eqref{H} can be created by driving the resonant carrier transition between the internal spin states using a microwave or radio-frequency field, which yields
\begin{equation}
\hat{H}_{y}(t)=\frac{\hbar\Omega_{y}(t)}{2}\left(e^{{\rm i}\phi}\left|\uparrow\right\rangle\left\langle\downarrow\right|+{\rm h.c}\right)=\frac{\hbar\Omega_{y}(t)}{2}\sigma_{y}.
\end{equation}
Here $\Omega_{y}(t)$ is the time-dependent Rabi frequency and we set the driving phase to $\phi=-\pi/2$. In the interaction picture rotating at the frequency $\omega$ the total Hamiltonian $\hat{H}=\hat{H}_{\rm s-b}+\hat{H}_{y}+\hat{H}_{F}$ is given by Eq. (\ref{H}) where the symmetry-breaking term is
\begin{equation}
\hat{H}_{\rm pert}=\frac{z_{0}F_{\rm d}}{2}(\hat{a}^{\dag}+\hat{a})\label{H_force}.
\end{equation}

The force sensing protocol starts by initialization of the spins along the $y$ direction and laser cooling of the single ion vibrational mode to the motional ground state. Subsequently, the transverse field exponentially decays as $\Omega_{y}(t)=\Omega_{y}(0)e^{-\gamma t}$, which drives the system adiabatically into the superposition state $\left|\psi_{\rm g}(t)\right\rangle=c_{+}(t)\left|\psi_{+}\right\rangle+c_{-}(t)\left|\psi_{-}\right\rangle$.
Here $c_{\pm}(t)$ are the respective probability amplitudes which are solutions of the time-dependent Schr\"odinger equation with the Hamiltonian (\ref{Heff1}) (see the Appendix).
In our scheme the detection of the force is performed either by measuring the expectation value of $\sigma_{x}$ or by measuring the position quadrature $\hat{Z}=\hat{a}^{\dag}+\hat{a}$ of the bosonic field.

\begin{figure}
\includegraphics[width=0.45\textwidth]{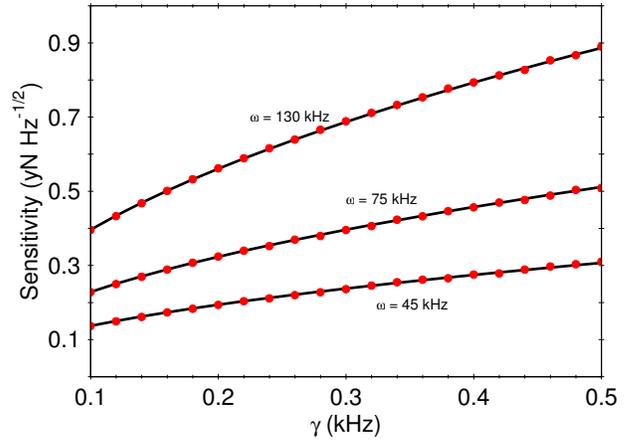}
\caption{(Color online) The sensitivity $\eta_{\rm force}\sqrt{T}$ versus the slope $\gamma$ for various values of $\omega$. We assume a single $^{24}$Mg$^+$ ion with an axial trap frequency $\omega_{z}=6.3$ MHz. The interaction time is set to $t_{f}=14\gamma^{-1}$. The other parameters are $g=25$ kHz, $\Omega_{y}(0)=225$ kHz. The exact solution (dots) is compared with the analytical expression $\eta_{\rm force}\sqrt{T}=\sqrt{t_{f}}F_{\rm d}^{\rm min}$ (solid line) where $F_{\rm d}^{\rm min}$ is given by Eq. (\ref{force_min}).}
\label{fig2}
\end{figure}

For vanishing force ($F_{\rm d}=0$) the parity symmetry is restored by creating an entangled ground state that is an equal superposition of states $\left|\psi_{\pm}\right\rangle$, which leads to $\langle \sigma_{x}(t_{f})\rangle=0$ and $\langle\hat{Z}(t_{f})\rangle=0$.
For $F_{\rm d}\neq0$, the parity symmetry of $\hat{H}_{\rm D}$ is broken, which allow us to estimate $F_{\rm d}$ by measuring $\sigma_{x}$ or $\hat{Z}$.
In Fig. \ref{fig1}(a) the time-evolution of the expectation value of $\sigma_{x}$ in the presence of the symmetry-breaking term \eqref{H_force} is shown.
At the interaction time $t_{f}\gg\gamma$ the signal and its variance are described by (see Appendix)
\begin{subequations}
\label{signal_Jx}
\begin{eqnarray}
&&\langle\sigma_{x}(t_{f})\rangle=\tanh\left(\frac{\pi g z_{0}F_{\rm d}}{\hbar\gamma\omega}\right),\label{s}\\
&&\langle\Delta^{2}\sigma_{x}(t_{f})\rangle=1-\langle \sigma_{x}(t_{f})\rangle^{2}.
\label{v}
\end{eqnarray}
\end{subequations}
Note that compared to other schemes here the sign of the force is fully preserved due to the tanh dependence of the signal.
The corresponding signal-to-noise ratio ${\rm SNR}=\langle \sigma_{x}(t_{f})\rangle/\langle\Delta^{2}\sigma_{x}(t_{f})\rangle^{1/2}$ is shown in Fig. \ref{fig1}(b).
The minimum detectable force is determined by the condition that the signal-to-noise ratio is equal to one, which gives
\begin{equation}
F_{\rm d}^{\rm min}=\frac{\hbar\gamma\omega}{\pi gz_{0}}\sinh^{-1}(1).\label{force_min}
\end{equation}
The sensitivity of the force measurement is defined as $\eta_{\rm force}=F_{\rm d}^{\rm min}/\sqrt{\nu}$, where $\nu=T/\tau$ is the repetition number with $T$ being the total experimental time.
The time $\tau$ includes the evolution time as well as the preparation and measurement times.
Because our scheme relies on the projective measurement of the spin populations at the end of the adiabatic transition we have $\tau\approx t_{f}$.
The sensitivity characterizes the minimal force difference, which can be discriminated within a total experimental time of one second.
In Fig. \ref{fig2} we show the sensitivity as a function of the slope $\gamma$ for various values of $\omega$.
Lowering $\gamma$ implies a longer interaction time $t_{f}$ and thus better sensitivity.
For example, using the parameters in Fig. \ref{fig2} for $\omega=45$ kHz and interaction time $t_{f}=30$ ms we estimate force sensitivity of about $0.3$ yN$/\sqrt{\rm{Hz}}$.
Further increasing of the sensitivity to $0.16$ yN$/\sqrt{\rm{Hz}}$ can be achieved with the interaction time of $t_{f}=100$ ms.

\subsec{Position quadrature}
Alternatively, the force estimation can be carried out by measuring the expectation value of the position quadrature $\langle \hat{Z}(t_{f})\rangle$. We find
\begin{eqnarray}
&&\langle \hat{Z}(t_{f})\rangle=-2\frac{g}{\omega}\tanh\left(\frac{\pi g z_{0}F_{\rm d}}{\hbar\gamma\omega}\right),\notag\\
&&\langle\Delta^{2}\hat{Z}(t_{f})\rangle=1+4\frac{g^{2}}{\omega^{2}}-\langle \hat{Z}(t_{f})\rangle^{2}.\label{singal_x}
\end{eqnarray}
Using Eq.~\eqref{singal_x} we obtain that for $\omega>2g$ the uncertainty of the position quadrature is higher than the measured signal ($\rm SNR<1$).
At $\omega=2g$ and in the limit $F_{\rm d}\gg 2\gamma\hbar/(\pi z_{0})$, the $\rm SNR$ tends asymptotically to one from below and thus no measurement is possible, whereas for $\omega<2g$ the force estimation is bounded by
\begin{equation}
F_{\rm d}^{\rm min}=\frac{\hbar\gamma\omega}{\pi g z_{0}}\tanh^{-1}\sqrt{\frac{1}{2}+\frac{\omega^{2}}{8g^{2}}}.\label{force_z}
\end{equation}
Comparing Eqs.~\eqref{force_min} and \eqref{force_z} we conclude that the signal of $\sigma_{x}$ provides better sensitivity.

\begin{figure}
\includegraphics[width=0.45\textwidth]{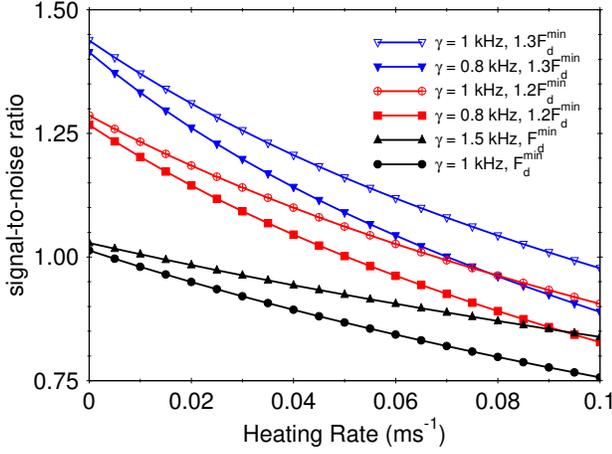}
\caption{(Color online) Signal-to-noise ratio versus the heating rate for various values of $\gamma$ and $F_{\rm d}$. We integrate numerically the master equation \eqref{ME} with the Hamiltonian \eqref{H}. The other parameters are set to $g=25$ kHz, $\omega=150$ kHz and $\Omega_{y}(0)=225$ kHz.}
\label{fig4}
\end{figure}

\subsec{Effect of motional heating}
The main source of decoherence which limits the force estimation in our scheme is caused by the motional heating.
In order to account for the effect of the motional heating in the sensing protocol we numerically integrate the master equation
\begin{eqnarray}
\frac{d \hat{\rho}}{dt}&=&-\frac{\rm{i}}{\hbar}[\hat{H},\hat{\rho}]+\frac{\gamma_{\rm dec}}{2}(\bar{n}+1)(2\hat{a}\hat{\rho}\hat{a}^{\dag}-\hat{a}^{\dag}\hat{a}\hat{\rho}-\hat{\rho}\hat{a}^{\dag}\hat{a})\notag\\
&&+\frac{\gamma_{\rm dec}}{2}\bar{n}(2\hat{a}^{\dag}\hat{\rho}\hat{a}-\hat{a}\hat{a}^{\dag}\hat{\rho}-\hat{\rho}\hat{a}\hat{a}^{\dag}).\label{ME}
\end{eqnarray}
Here $\gamma_{\rm dec}$ is the system decay rate and $\bar{n}$ is the average number of quanta in the reservoir. In the limit $\bar{n}\gg1$ the motional heating is characterized by the time $t_{\rm dec}=1/(\bar{n}\gamma_{\rm dec})$ and the heating rate is $\langle \dot{n}\rangle=1/t_{\rm dec}$. In Fig. \ref{fig4} we show the signal-to-noise ratio as a function of $\langle \dot{n}\rangle$.
By increasing the heating rate, the  corresponding signal-to-noise ratio decreases with stronger damping for lower $\gamma$ and $\omega$.
Note that for small $\gamma$ the effect of the motional heating is stronger, since the evolution time $t\sim \gamma^{-1}$ is longer.
Using the parameters presented in Fig. \ref{fig4} we estimate sensitivity of $1.9$ yN$/\sqrt{\rm{Hz}}$ within the interaction time $t_{f}=14$ ms and heating rate approximately of $0.1$ $\rm{ms}^{-1}$, which corresponds to typical heating rates in linear ion Paul traps. For a cryogenic ion trap with heating rate of order of $\langle \dot{n}\rangle=0.01$ $\rm{ms}^{-1}$ \cite{Chiaverini2014} the force sensitivity would be approximately of $0.7$ yN$/\sqrt{\rm{Hz}}$ for $\omega=75$ kHz and $\gamma=0.8$ kHz.

\sec{Sensing of spin-dependent forces}
Here we introduce a measurement protocol that can be used for sensing of spin-dependent forces.
The origin of such forces can be due to (i) spatially dependent Zeeman shifts experiences by spins in a magnetic-field gradient \cite{Mintert2001} or (ii) spatially dependent Stark shifts experienced by atoms in an intensity field gradient, which give rise to an optical dipole force \cite{Hume2011}. Consider the case of an oscillating magnetic field gradient along the trap axis. Consider also an ion chain with two ions where the internal states of the first ion are formed by the magnetic insensitive clock states $|s\rangle_{\rm cl}$ with $s=\uparrow,\downarrow$. These clock states are used to implement the quantum Rabi Hamiltonian (\ref{Dicke}) taking $\sigma_{x}=(\left|\uparrow\right\rangle_{\rm cl}\left\langle\downarrow\right|+{\rm h.c.})$ and $\sigma_{y}={\rm i}(\left|\downarrow\right\rangle_{\rm cl}\left\langle\uparrow\right|-{\rm h.c.})$ with the same method as was described above. The auxiliary ion (not necessarily the same atomic species) is prepared in one of its magnetic sensitive states $|\rm{aux}\rangle$. The external oscillating magnetic field gradient $\vec{B}(t)=\cos(\omega t)B_{0} z \vec{e}_{z}$ will create a coupling between the auxiliary ion states and the collective vibrational modes. Assuming that the oscillating frequency is close to the center-of-mass vibrational mode $\omega=\omega_{\rm c.m.}-\omega$ and neglecting the fast oscillating terms we arrive to
\begin{equation}
\hat{H}_{\rm pert}=\frac{z_{\rm c.m.}F_{z}}{\sqrt{2}}(\hat{a}^{\dag}+\hat{a})|{\rm aux}\rangle\langle{\rm{aux}}|,\label{H_mg}
\end{equation}
where $z_{
\rm c.m.}=\sqrt{\hbar/2 m\omega_{\rm c.m.}}$. Here
$F_{z}=g_{J}\mu_{\rm B} B_{0}/2$ is the force acting on the magnetic moment $\mu_{z}^{\rm aux}=(g_{J}\mu_{\rm B}/2)|\rm{aux}\rangle\langle\rm{aux}|$ of the auxiliary ions where $g_{J}$ is the Lande g-factor and $\mu_{\rm B}$ is the Bohr magneton.
The force acting on the clock ion depends on the spin state of the auxiliary ion such that the perturbation term becomes
\begin{equation}
\left\langle\rm{aux}\right|\hat{H}_{\rm pert}\left|\rm{aux}\right\rangle=\frac{z_{\rm c.m.}F_{z}}{\sqrt{2}}(\hat{a}^{\dag}+\hat{a}).
\end{equation}
The induced force on the clock ion can be used for studying, e.g., the magnetic moment of the auxiliary atomic or molecular ions trapped together with the clock ion. Indeed, assuming magnetic dipole moment $2\mu_{\rm B}$ of the auxiliary ion and magnetic field gradient of $1$ T/m the resulting force is about $9$ $\rm{yN}$, which can be detected by measuring the spin population of the clock ions.

\sec{Conclusions}\label{C}
We have shown that the system described by the quantum Rabi Hamiltonian can serve as a detector of extremely small forces.
The underlying physical mechanism is the process of symmetry-breaking adiabatic transition due to the presence of force perturbations.
Our sensing protocols can be implemented using a trapped ion, where the parameters which drive the system across the adiabatic transition are controlled by external laser or microwave fields.
We have shown that a system of a single trapped ion can be used as a probe for electric sensing with sensitivity about and even below 1 yN$/\sqrt{\rm{Hz}}$ range. Additionally, the proposed method can be extended for sensing magnetic fields.
{A major advantage of our protocol is that it demands a single population measurement, thereby achieving a considerably speed-up over previous protocols using phonon number measurement via Rabi oscillations.}

%%%%%%%%%%%%%%%%%%%%%%%%%%%%%%%%%%%%%%%%%%%%%%%%%%%%%%%%%%%%%%%%%%%%%%%%%%%%%%%%%%%%%%%%%%%%%%%%%%%%%%%%%%%%%%%%%%%%%%%%%%%%%%%%%%%
\acknowledgments

This work has been supported by the EC Seventh Framework Programme under Grant Agreement No. 270843 (iQIT).

%\appendix
\section{Appendix: Derivation of the signal and the variance of the signal}
In the symmetry-broken phase the underlying dynamics of the QM Hamiltonian with the perturbation term \eqref{H_force} can be captured within the two-level model with the effective Hamiltonian
\begin{equation}
H_{\rm eff} =\left[
\begin{array}{cc}
- z_{0}F_{\rm d}\frac{g}{\omega} & \hbar\Delta_{\rm gap}(t)/2 \\
\hbar\Delta_{\rm gap}(t)/2 &  z_{0}F_{\rm d}\frac{g}{\omega}   \\%
\end{array}%
\right].\label{Heff}
\end{equation}
The probability amplitudes $c_{\pm}(t)$ for the state vector $\left|\psi_{\rm g}(t)\right\rangle=c_{+}(t)\left|\psi_{+}\right\rangle+c_{-}(t)\left|\psi_{-}\right\rangle$ obey the time-dependent Schr\"odinger equation, which turns into a system of two coupled differential equations,
\bse
\begin{eqnarray}
&&{\rm i}\dot{c}_{+}(t)=-\frac{z_{0}F_{\rm d}g}{\hbar \omega}c_{+}(t)+\frac{\Delta_{\rm gap}(t)}{2}c_{-}(t),\\
&&{\rm i}\dot{c}_{-}(t)=\frac{z_{0}F_{\rm d}g}{\hbar \omega}c_{-}(t)+\frac{\Delta_{\rm gap}(t)}{2}c_{+}(t),
\end{eqnarray}
\ese
subject to the initial conditions $c_{+}(t_{i})=1/\sqrt{2}$ and $c_{-}(t_{i})=-1/\sqrt{2}$. We assume that the transverse field varies in time as $\Omega_{y}(t)=\Omega_{y}(0)e^{-\gamma t/N}$, which implies that $\Delta_{\rm gap}(t)=\Delta_{i}e^{-\gamma t}$ and the two-state problem reduces to the Demkov model. The probabilities at $t_{f}\gg\gamma^{-1}$ are given by
\begin{equation}
|c_{+}(t_{f})|^{2}=\frac{1}{2}+\frac{1}{2}\tanh\left(\frac{\pi g z_{0}F_{\rm d}}{\hbar\gamma\omega}\right)
\end{equation}
and $|c_{-}(t_{f})|^{2}=1-|c_{+}(t_{f})|^{2}$. The expectation value and the variance of $\sigma_{x}$ with respect to $\left|\psi(t_{f})\right\rangle$ are
\bse
\begin{eqnarray}
&&\langle \sigma_{x}(t_{f})\rangle=2|c_{+}(t_{f})|^{2}-1,\\
&&\langle\Delta^{2}\sigma(t_{f})\rangle=4|c_{+}(t_{f})|^{2}|c_{-}(t_{f})|^{2}.\label{sig_var}
\end{eqnarray}
\ese
For the measured signal of the quadrature $\hat{Z}$ and its variance we find
\bse
\begin{eqnarray}
&&\langle Z(t_{f})\rangle=2\alpha\left(2|c_{+}(t_{f})|^{2}-1\right),\\
&&\langle\Delta^{2} Z(t_{f})\rangle=1+16\alpha^{2}|c_{+}(t_{f})|^{2}|c_{-}(t_{f})|^{2}.
\end{eqnarray}
\ese
%%%%%%%%%%%%%%%%%%%%%%%%%%%%%%%%%%%%%%%%%%%%%%%%%%%%%%%%%%%%%%%%%%%%%%%%%%%%%%%%%%%%%%%%%%%%%%%%%%%%%%
%%%%%%%%%%%%%%%%%%%%%%%%%%%%%%%%%%%%%%%%%%%%%%%%%%%%%%%%%%%%%%%%%%%%%%%%%%%%%%%%%%%%%%%%%%%%%%%%%%%%%%

\end{document}